\documentclass[twocolumn,prb,aps,amsmath,showpacs]{revtex4}
\usepackage{graphicx}
\usepackage[version=3]{mhchem}
\usepackage{dcolumn}
\usepackage{multirow}
\usepackage{bm}
\usepackage{subfigure}
\usepackage{amssymb}
\usepackage{color}
\begin{document}

\title{Structural, electronic and optical properties of well-known primary explosive: Mercury Fulminate}
\author{N. Yedukondalu and G. Vaitheeswaran$^*$ }
\affiliation{Advanced Centre of Research in High Energy Materials (ACRHEM),
University of Hyderabad, Prof. C. R. Rao Road, Gachibowli, Hyderabad- 500046, Telangana, India.}
\date{\today}

\begin{abstract}
Mercury Fulminate (MF) is one of the well-known primary explosives since 17$^{th}$ century and it has rendered invaluable service over many years. However, the correct molecular and crystal structures are determined recently after 300 years of its discovery. In the present study, we report pressure dependent structural, elastic, electronic and optical properties of MF. Non-local correction methods have been employed to capture the weak van der Waals interactions in layered and molecular energetic MF. Among the non-local correction methods tested, optB88-vdW method works well for the investigated compound. The obtained equilibrium bulk modulus reveals that MF is softer than the well known primary explosives Silver Fulminate (SF), silver azide and lead azide. MF exhibits anisotropic compressibility (b$\textgreater$a$\textgreater$c) under pressure, consequently the corresponding elastic moduli decrease in the following order: C$_{22}$$\textgreater$C$_{11}$$\textgreater$C$_{33}$. The structural and mechanical properties suggest that MF is more sensitive to detonate along c-axis (similar to RDX) due to high compressibility of Hg...O non-bonded interactions along that axis. Electronic structure and optical properties were calculated including spin-orbit (SO) interactions using full potential linearized augmented plane wave method within recently developed Tran-Blaha modified Becke-Johnson (TB-mBJ) potential. The calculated TB-mBJ electronic structures of SF and MF show that these compounds are indirect band gap insulators. Also SO coupling is found to be more pronounced for $4d$ and $5d$-states of Ag and Hg atoms of SF and MF, respectively. Partial density of states and electron charge density maps were used to describe the nature of chemical bonding. Ag-C bond is more directional than Hg-C bond which makes SF to be more unstable than MF. The effect of SO coupling on optical properties has also been studied and found to be significant for both (SF and MF) of the compounds. 
\end{abstract}
\maketitle
\section {Introduction}
Based upon sensitivity to the external stimuli such as heat, shock, impact, friction, or electric charge, the high explosives are classified into two categories namely primary and secondary. A typical explosive consists of a main charge of secondary explosive with a high output but low sensitivity to initiation, which is initiated by an adjacent primary explosive, which transmits a sufficiently strong shock to the secondary explosive which leads to detonation.\cite{dickson} Inorganic fulminates come under the class of primary explosives, they find applications as initiators for secondary explosives and are iso-electronic with the corresponding azides, cyanates and cyanamides.\cite{iqbal1} MF was the first, widely and long been used as primary explosive due to its excellent priming power, high performance, and it can be easily detonated.\cite{klapotke} MF has rendered invaluable service over many years and this can be clearly seen from its annual production only in Germany was about 1,00,000 kg per year in the beginning of 20$^{th}$ century.\cite{kurzer} The wide application of dynamite was only possible when the use of MF as initiator which guarantees a safe ignition and hence it is used to initiate dynamite in metal blasting cap detonator.\cite{kurzer,nobel} MF detonate after the initiation with external stimuli by producing CO, N$_2$, and Hg as the decomposition products: Hg(CNO)$_2$ $\rightarrow$ Hg + 2CO + N$_2$.\cite{berthelot} MF is very sensitive to shock, impact, friction and sunlight. MF is detonated by sparks and flames\cite{klapotke} and also it is desensitized by addition of water. Since LA was found to detonate more reliably (detonation velocity 4.25 km/s for MF and 5.3 km/s for LA), less impact sensitivity (1-2 N m for MF and 2.5-4 N m for LA) and to have better thermal stability (temperature of ignition 210$^o$C for MF and above 300$^o$C for LA) and hence MF was largely replaced by LA.\cite{miles1,collins,meyer}

\par Several methods were proposed in the literature to synthesis MF, among them, Howard's interpretation for the formation of MF from mercury, nitric acid and ethanol was widely accepted.\cite{wieland1,wieland2,wohler} However, MF has been used as a primary explosive for a long time but the determination of its crystal symmetry is an extensive debate until 2007. Since 1931 several investigations have been made to determine the crystal structure of MF using single crystal\cite{miles2, suzuki,iqbal1} and powder X-ray diffraction methods\cite{ICDD,brown} but these attempts were unsuccessful to determine correct crystal structure of MF. Recently, Beck et al\cite{beck} reported the correct crystal structure of this energetic compound. Moreover, Density Functional Theory (DFT) calculations have been carried out for a single molecule of MF at B3LYP level and they predicted bent CNO-Hg-ONC units of molecular structure.\cite{lemi} Once again Beck et al\cite{klapotke} made a detailed theoretical investigation on molecular structure of MF and they proved that the molecular and Lewis structure of MF is linear in gas phase $i.e.$ ONC-Hg-CNO. The molecular structure is in contrast to the previous theoretical prediction\cite{lemi} but it is in good accord with their recent X-ray diffraction study.\cite{beck} In addition the authors also proposed that Hg-C-N angle is 180 $^o$ in isolated molecule whereas it is 169 $^o$ in the crystalline solid form which is due to intermolecular interactions and packing effects.\cite{klapotke} However, except the crystal structure most of the fundamental physical properties are unknown for the investigated compound at electronic level. With this motivation, we performed a detailed study of structural and mechanical properties under pressure up to 5 GPa using advanced dispersion corrected methods and electronic structure, optical properties by including SO interactions at ambient pressure by means of first principles calculations within the frame work of density functional theory (DFT). The rest of the article is organized as, in section II, we briefly describe methodology of our calculation. In section III, the structural, elastic, electronic structure and optical properties of MF are discussed. Finally, in section IV, we summarize the results, which concludes our paper.

\section{Computational details}
First principles calculations were performed using the Vienna $\emph{ab-initio}$ Simulation Package (VASP)\cite{Kresse} based on DFT with the all-electron projected augmented wave (PAW) method. The ion-cores are described within the PAW method while electron-electron interactions are treated with the Perdew-Burke-Ernzerhof (PBE)\cite{burke} parametrization of the generalized gradient approximation (GGA) with plane wave cut-off energy of 1000 eV and a 9$\times$9$\times$5 k-mesh according to the Monkhorst-Pack grid scheme.\cite{monkhorst} Quasi-Newton algorithm is used to relax the ions and the system was fully relaxed with residual forces smaller than 0.001 eV/$\AA$.
\par In order to treat weak dispersive interactions, there are two kinds of dispersion corrections; first one is pairwise additive correction, second one is non-local correction and both of these methods have shown remarkable success recently. In the first method, vdW parameters for heavy metals (namely Cs, Ba, Hg, Tl, Pb and Bi etc.) $i.e.$ 6$^{th}$ and 7$^{th}$ periods of the periodic table elements are not well optimized whereas the second method is used to study the simple as well as heavy metal based systems and the results show success of this method in treating the van der Waals (vdW) interactions for wide range of materials.\cite{} Therefore, in the present study, we have used the second method so-called non-local correction method proposed by Dion et al\cite{Dion} and further modified by Klimes et al,\cite{Klimes} in which the vdW contribution to the total energy is described through modifications to the correlation energy functional within DFT. Specifically, the DFT exchange-correlation functional takes the form:
\begin{center}
E$_{xc}$ = E$_x^{GGA}$ + E$_c^{LDA}$ + E$_c^{nl}$ \\
\end{center}  Here E$_x^{GGA}$ is the exchange energy,\cite{burke} E$_c^{LDA}$ is the local density approximation (LDA) correlation energy\cite{perdew} and E$_c^{nl}$ is the non-local correction which is given by
\begin{center}
E$_c^{nl}$ = $\frac{1}{2}$ $\displaystyle{\int}$ dr $\displaystyle{\int}$ n(r)$\phi$(r,r$'$)n(r$'$) dr$'$ \\
\end{center}
where n(r) is the electron density and the kernel $\phi$(r,r$'$) is a function of n(r) and n(r$'$), their gradients, and r-r$'$. However, this method requires massive computation to evaluate the double integral in the above equation using the fast-Fourier transform grid points, especially for large cells.\cite{vashishta}

It is well known fact that the standard DFT functionals severely underestimate the band gap by 30-40$\%$ for semiconductors and insulators.\cite{byrd} In contrast to LDA/GGA functionals, recently developed Tran-Blaha modified Becke Johnson (TB-mBJ)\cite{peter} potential shows remarkable success in predicting the energy band gaps for diverse materials\cite{singh,camargo,dixit,jiang} and competing with the computationally expensive methods such as GW approximation and hybrid functionals. Therefore, in the present work, TB-mBJ potential has been used to get reliable energy band gap thereby calculation of electronic structure and optical properties of SF and MF. This semi-local potential is implemented through WIEN2K package.\cite{blaha} To achieve the required convergence of energy eigenvalues, the wave functions in the interstitial region were expanded using plane waves with a cut-off K$_{max}$ = 7/RMT while the charge density was Fourier expanded up to G$_{max}$ = 14, where Radius of Muffin Tin (RMT) is the smallest atomic sphere radius and K$_{max}$ denotes the magnitude of the largest K vector in plane wave expansion. The RMT radii are assumed to be 2.0, 1.05, 1.05 and 1.25 Bohrs for Ag/Hg, C, N and O, respectively. The wave functions inside the spheres are expanded up to $l_{max}$ = 10. Self-consistency of total energy is obtained by using 9$\times$9$\times$5 k-mesh in the Irreducible Brillouin Zone (IBZ). The frequency-dependent optical properties have been calculated using a denser k-mesh of 19$\times$ 19$\times$12 in the IBZ.

\section{Results and discussion}
\subsection{Crystal structure}
MF is a long standing primary explosive but the molecular geometry and crystal structure of MF has been resolved more than 300 years of after its discovery.\cite{beck} The chemical formula of MF $i.e.$ Hg(CNO)$_2$ is analogous to the corresponding Mercury Azide (MA), Hg(NNN)$_2$. Moreover, the fulminate and/or azide single anion is linear and contain 16 valence electrons resulting a negative charge. MF crystallizes in the orthorhombic centro symmetric space group \emph{Cmce} with lattice parameters a = 5.470$\AA$, b = 10.376$\AA$, c = 7.70$\AA$, V = 437.03 $\AA^3$, and Z = 4.\cite{beck} While MA crystallizes in non-centro symmetric space group \emph{Pca2$_1$} with lattice parameters a = 10.632$\AA$, b = 6.264$\AA$, c = 6.323$\AA$, V = 421.10 $\AA^3$ and Z = 4,\cite{muller} consequently MA is $\sim$ 4$\%$ more densely packed than MF.\cite{beck} As shown in figure 1, the crystal structure consists of MF molecule at each corner as well as face centre of the unit cell (see figure 1a), the planar MF molecules are located at x = 0 and x = 0.5 along b-axis\cite{beck} and the layers are stacked along a and c-axes as depicted in figure 1b and 1c, respectively. Apart from this, experimental measurements reveal that the arrangement of MF molecules in b-c plane leads to two non-bonded contacts between Hg and O atoms (see figure 1d) with a distance of Hg...O = 2.833 $\AA$ within the unit cell, which is less than the sum of the vdW radii 3 $\AA$ of Hg and O atoms (vdW radii 1.5 $\AA$ for Hg and O atoms) which causes weak vdW interactions in the crystalline MF.\cite{beck} The intermolecular interactions play a significant role in predicting the structure and stability of the layered and molecular crystalline solids.

The effect of SO coupling is of minor importance for structural optimization.\cite{Wang,Shein,kanchana,Albanesi} Therefore, we first obtain the ground state crystal structure of MF by performing full structural optimization of both lattice constants and internal co-ordinates without inclusion of SO interactions. The obtained equilibrium volume of MF is overestimated by $\sim$ 20.9$\%$ within PBE-GGA functional. This clearly represents that the standard PBE-GGA functional is inadequate to predict the ground state properties of the energetic layered and molecular solid MF. Recently, usage of non-local correction methods become successful in describing the structural properties of energetic molecular solids,\cite{landerville,Kondaiah,vashishta,Sorescu1} nitrogen rich salts,\cite{Sorescu2} organic-inorganic hybrid perovskite,\cite{Proupin,Wang} and layered materials.\cite{Graziano} With this motivation, we have also used various non-empirical dispersion corrected methods to capture vdW interactions to reproduce the ground state properties which are comparable with the experiment.\cite{beck} The computed ground state volume with non-local dispersion corrected methods for MF is overestimated by around 1.7$\%$ using vdW-DF; 5.9$\%$ using vdW-DF2 and underestimated by around 0.4$\%$ using optB88-vdW; 1.1$\%$ using optB86b-vdW methods. Among the examined non-local dispersion corrected methods, optB88-vdW method works well for the MF. The small discrepancies between theoretical values at 0 K and experimental data at 295 K\cite{beck} were observed. The order of discrepancies about $\sim$0.4-5.9$\%$ are previously reported for secondary explosive molecular crystals with vdW-DF methods at 0 K.\cite{landerville,vashishta,Sorescu1} The calculated ground state unit cell lattice constants, volume and density of MF using various non-local correction methods are compared with the experimental data\cite{beck} and are presented in Table I. In addition, we have also calculated the intra-molecular interactions for equilibrium structure obtained using optB88-vdW method. The calculated bond lengths Hg-C, C-N and N-O are 2.028 (2.029), 1.172 (1.143), 1.235 (1.248) $\AA$ and bond angles Hg-C-N, C-N-O and C-Hg-C are 167.7 (169.1), 179.7 (179.7) and 180 (180) $^o$ respectively, which are in good agreement with the experimental\cite{beck} results given in parenthesis. The C$\equiv$N bond length in an isolated molecule is 1.160 $\AA$ and the calculated value for MF is 1.172 $\AA$ which is strongly suggesting that there exists a triple bond between C and N atoms as observed in the experiment (d$_{(C\equiv N)}$ = 1.143 $\AA$).\cite{beck} As discussed in section I, MF possesses perfect linear molecular structure in single molecular gas phase.\cite{klapotke} While in the crystalline solid form, the calculated angle between Hg-C and C-N bonds is 167.7$^o$ and it is deviated by 12.3$^o$ from linearity (180$^o$) which is in good agreement with experimental\cite{beck} deviation of 11$^o$. This deviation clearly indicates that slightly distorted linear molecular structure of MF in the crystalline form when compared to its molecular structure in gas phase.

\subsection{Equation of State and Compressibility}
We turned our attention to investigate the effect of hydrostatic pressure on crystal structure of MF. In order to understand the behavior of unit-cell parameters and their relative compressibilities under compression, we have presented the lattice constants as a function of pressure. The pressure dependent lattice constants show that a and c lattice constants decrease whereas lattice constant 'b' increases with pressure. Increase of lattice constant 'b' under hydrostatic pressure is interesting in MF and this is similar to the case of silver azide (SA)\cite{zhu} in which lattice constant 'a' increases as a function of pressure for ambient phase ($Ibam$). This clearly indicates the anisotropic behavior of lattice constants under the studied pressure range as depicted in figure \ref{abc_MF}a. Equation of state (EOS) represents the functional relationship between the thermodynamic variables (pressure, volume and temperature) for solids. The calculated volume decrease monotonically as a function of pressure as shown in figure \ref{abc_MF}b. By fitting pressure-volume data to third-order Birch-Murnaghan equation of state,\cite{murnaghan} the obtained equilibrium bulk modulus (B$_0$) and its pressure derivative are found to be 12.2 GPa and 7.9 respectively using optB88-vdW. However, the calculated B$_0$ value 12.2 GPa for MF is lower than that of SF (20 GPa),\cite{kondal} SA (39 GPa)\cite{Hou} and LA (26 GPa\cite{millar} and 41 GPa\cite{weir}) which indicates the soft nature of MF when compared to other well-known primary explosives. 

Further, to understand the compressibility of MF, normalized lattice constants, bond lengths and bond angles are plotted as a function of pressure as displayed in figure \ref{BLA_MF}. The pressure dependent lattice constants show anisotropic axial compressibilities of 96.2$\%$, 102.4$\%$ 83.3$\%$ along a, b, and c crystallographic directions, respectively and the order of compressibility is as follows b$\textgreater$a$\textgreater$c. As depicted in figure \ref{BLA_MF}a, c-axis is the most compressible for MF which is due to high compressibility of non-bonded Hg...O intermolecular interactions along the c-axis as shown in figure \ref{BLA_MF}b. While the intra-molecular bonds Hg-C, C-N, and N-O show very less compressible nature (see figure \ref{BLA_MF}b) over the studied pressure range. This clearly shows that the intermolecular interactions are weaker than intra-molecular interactions in the layered MF. The bond angle Hg-C-N shows more compressible behavior whereas C-N-O and C-Hg-C exhibit less compressible nature under the studied pressure range as depicted in figure \ref{BLA_MF}c. Overall, we observe that Hg-C, C-N, and N-O bonds are stiffer whereas Hg...O non-bonded distance is more compressible under the application of hydrostatic pressure. 

\subsection{Elastic constants and mechanical properties}
Elasticity describes the response of a crystal under external strain which gives an information about the bonding characteristics for the anisotropic character of the solid.\cite{Dong} Quantifying and understanding the elastic response of energetic materials is a necessary first step towards determining the mechanical and chemical mechanisms that produce this anisotropic behavior under shock loading.\cite{Winey} Numerous researchers focused their attention on understanding detonation initiation by mechanical shock.\cite{Haycraft1} Detonation of an energetic material can be considered as a collective property of the material and is highly depends upon intermolecular interactions, molecular arrangements, and molecular composition which has a measurable effect on the macroscopic properties of the energetic solid.\cite{Haycraft2} Therefore, we focused our attention to understand the elastic behavior of energetic MF. Due to orthorhombic crystal symmetry, MF has nine independent elastic constants namely C$_{11}$, C$_{22}$, C$_{33}$, C$_{44}$, C$_{55}$, C$_{66}$, C$_{12}$, C$_{13}$, and C$_{23}$. As presented in Table II, the calculated elastic constants are positive and obey the Born's mechanical stability criteria,\cite{Born} which indicate that MF is mechanically stable at ambient pressure. A direction in which intermolecular interactions are weak would reflect a higher compressibility along that direction. The compressibility of orthorhombic lattice constants a, b, and c can be directly correlated with the diagonal elastic constants C$_{11}$, C$_{22}$, and C$_{33}$, respectively. As discussed in the above section, the compressibility order for the investigated compound is b$\textgreater$a$\textgreater$c which reveals that MF has the weakest interactions along the c-axis due to weak intermolecular interactions along c-axis (see figure 1d). Consequently, C$_{33}$ possesses lowest value in magnitude among the three diagonal elastic moduli and they decrease in the following order C$_{22}$$\textgreater$C$_{11}$$\textgreater$C$_{33}$ as compressibility order of the lattice constants (b$\textgreater$a$\textgreater$c). Previously Haycraft et al\cite{Haycraft1,Haycraft2} made a correlation between linear compressibility and elastic constants thereby relevance to shock detonation sensitivity for RDX and CL-20 single crystals. They reported that RDX and CL-20 are found to be more sensitive to detonation along c and a-axes, respectively. On the similar path, the calculated compressibility and elastic moduli disclose that MF is found to be more sensitive to detonation along the c-axis. The other three diagonal elastic constants decrease as follows: C$_{55}$$\textgreater$C$_{44}$$\textgreater$C$_{66}$. C$_{66}$ and C$_{44}$ are found to be relatively small compared to C$_{55}$, which is an indication of the soft shear transformation along (001) and/or (100) planes. On the other hand, three off-diagonal elastic constants (C$_{12}$, C$_{13}$, and C$_{23}$); C$_{12}$ and C$_{13}$ couple an applied normal stress component in the {\bf a}-direction with uniaxial strain along {\bf b}- and {\bf c}-axis respectively while C$_{23}$ couples a applied normal stress along {\bf b}-direction with an uniaxial strain along {\bf c}-axis.\cite{Haycraft2} C$_{23}$ has the largest value among the three transverse coupling elastic moduli and the low values of C$_{12}$ and C$_{13}$ would suggest that the crystal system is susceptible to shear along the crystallographic {\bf b}- and {\bf c}-axes when normal stress is applied along crystallographic {\bf a} direction. In addition, we have also calculated the elastic moduli as a function of pressure. As depicted in figure \ref{Cij_MF}, all the elastic moduli increase (except C$_{66}$), especially C$_{22}$ grows rapidly as a function of pressure. However, we observe a softening of C$_{66}$ elastic constant with pressure which may induce shear instability in MF under high pressure. 

When mono-crystalline samples are not available then it is not possible to measure the single crystal elastic constants. Instead, the polycrystalline bulk and shear moduli may be determined $i.e.$ the average isotropic elastic moduli can be obtained from anisotropic single crystal elastic moduli.\cite{Ravindran} The Vigot, Reuss and Hill approximations can predict the theoretical maximum, minimum and average polycrystalline elastic moduli, respectively. The obtained B$_0$ value 12.2 GPa from EOS is comparable with the derived B$_R$ value of 14.2 GPa. Shear modulus G$_R$ value 3.6 GPa is closely comparable (in magnitude) with the novel secondary explosive CL-20\cite{Haycraft1} using Reuss approximation and the low value of shear moduli indicates that overall MF is more susceptible to shear forces. In addition, we also made an attempt to calculate the sound wave velocities thereby Debye temperature of MF using the expressions given in Ref.\onlinecite{Kanchana} as presented in Table II using the isotrpic elastic moduli obtained from Hill approximation. Overall, the calculated polycrystalline elastic moduli, sound wave velocities and Debye temperature of MF are lower than the layered nitrogen rich alkali and alkaline-earth metal azide salts.\cite{ramesh} Furthermore, the stiffness of lattice and bond parameters can be clearly understand by analyzing the nature of chemical bonding for the investigated compound.

\subsection{Electronic structure and chemical bonding}
Silver and Mercury fulminates are iso-electronic with the corresponding azides, cyanates, and cynamides. Iqbal et al\cite{iqbal1} accomplished a detailed study on electronic structure and stability of inorganic fulminates, which reveals that nature of the bond between metal and carbon atoms is ionic in sodium, potassium and thallous fulminates whereas it is covalent in silver and mercury fulminate salts and this will be further reflected in their order of stability. Iqbal et al\cite{iqbal2} also proposed that the heavy metal based salts are unstable than light metal salts because of the asymmetric inter ionic distances. In addition, X-ray electron spectroscopy study\cite{Colton} on inorganic azides reveals that heavy metal azides are unstable than alkali metal azides due to their directional bonding nature. Therefore, the investigation of electronic structure and chemical bonding is vital to understand the stability of the energetic materials. 

From theoretical perspective, electronic structure calculations for silver and mercury fulminate salts are lacking in the literature. Since SO plays a significant role for heavy metals, in the present work, we have attempted a comparative analysis of electronic structure between SF and MF including SO interactions. In analogy to the Zeeman effect, when an electron moves in an electric field E, it experiences a magnetic field B$_{eff}$ $\sim$ E $\times$ $\frac{p}{mc^2}$ in its rest-frame (where m, p and c are mass, momentum of an electron and speed of light, respectively)-a field that induces a momentum-dependent Zeeman energy called the SO coupling, $\hat{H}_{SO}$ $\sim$ $\mu_B$ (E $\times$ p)·$\sigma$/mc$^2$, where $\sigma$ is the vector of the Pauli spin matrices and $\mu_B$ (= 9.27 $\times$ 10$^{−24}$ JT$^{-1}$) is the Bohr magneton. In crystals, the electric field is given by the gradient of the crystal potential E = -$\nabla$V, which produces a SO field w(p) = -$\mu_B$($\nabla$V $\times$ p)/mc$^2$.\cite{Manchon} We first optimized the fractional co-ordinates of both SF and MF at the experimental lattice constants\cite{Barrick,beck} within PBE-GGA using FP-LAPW method and are presented in Table III. The calculated band gaps are found to be 2.13 and 3.64 eV for SF and MF respectively at the PBE-GGA level. The PBE-GGA band gap value is slightly higher than LDA value of 2.0 eV\cite{kondal} for SF. The obtained TB-mBJ band gap values for SF and MF are found to be 3.32 and 4.92 eV respectively. When SO is included, the TB-mBJ band gaps are found to be 3.30 and 4.82 eV for SF and MF, respectively and the corresponding reduction in the band gap values after inclusion of SO are 0.02 and 0.1 eV. The small reduction in the band gap values are due to occurrence of SO splitting at the lower part of the valence band (VB) for SF (between -2 to -4 eV) and for MF (between -4.5 to -7 eV). The obtained TB-mBJ band gap values with SO are lowered by 0.7 eV for SF and increased by 0.42 eV for MF when compared to the optical energy gap measurements\cite{iqbal1} of 4.0 and 4.4 eV for SF and MF, respectively. However, wrong space group has been used for MF (which results a bent molecular structure of MF, which is in contrast to the recent experimental measurements.\cite{beck}) in the optical and spectroscopic measurements.\cite{suzuki,iqbal1} We have calculated band structures of both the compounds without and with SO coupling as presented in figures \ref{BS}a (\ref{BS}d) and \ref{BS}b (\ref{BS}e) for SF (MF), respectively. For clear understanding, we have also plotted the band structures without and with SO on top of each other as displayed in figures \ref{BS}c $\&$ f. As illustrated in figure \ref{BS}, SF and MF are indirect band gap insulators along S-($\Gamma$-Z) and R-$\Gamma$ directions, respectively. To a large extent the band structures of both the compounds look essentially similar with and without SO except for few bands in the lower part of the VB which are split due to the SO interactions as shown in the figures \ref{BS}c $\&$ f. There are few energetically low lying bands, three for SF and five for MF (see figure 1 of the supplementary material\cite{support}). In case of SF, the lowest bands in the VB region are derived from $s$-states of C, $p$-states of O atoms and the bands are positioned around -10 eV. The middle bands are due to $s,p$-states of fulminate group and finally the top of VB is mainly dominated by $p$-states of fulminate group, $d$-states of Ag atom and SO splitting is mainly due to $4d$-states of Ag atom. Especially, the bands along the high symmetry directions between U to R are split due to SO coupling for SF. While in case of MF, the lowest lying bands around -10 eV are due to $d$-states of Hg and $s$-states of C atoms and the $s,p$-states of fulminate group are positioned around -7.8 eV. The middle bands are derived from $s,p$-states of C, N, O atoms and $5d$-states of Hg atom, which are split due to SO in the energy range between -4.5 to -7.0 eV and the similar kind of splitting is seen for $5d$-bands of Hg atom in case of red-HgI$_2$.\cite{Ahuja} The bands around -2.5 eV are dominated by $d$, $s$-states of Hg atom and finally the top valence bands are mainly due to $p$-states fulminate group. From the calculated electronic band structures with and without SO, it is found that inclusion of SO is more significant for $4d$ and $5d$-bands of Ag and Hg atoms in the energy range -2 to -4 eV and -4.5 to -7 eV for SF and MF, respectively as depicted in figure \ref{BS}c $\&$ f.

Further, the intrinsic characteristics of chemical bonding in SF and MF was investigated by examining the total and partial density of states (PDOS). We have plotted the PDOS of SF and MF with and without SO as depicted in figure \ref{DOS}. As illustrated in figure \ref{DOS}, the conduction band is mainly due to $p$-states of C, N, O and $s,p,d$-states of metal (Ag/Hg) atoms. The lowest lying states positioned between -4.5 to -7.0 eV are due to hybridization of predominantly $5d$-sates of Hg which are split due to SO and anionic $p$-sates of C, N and O atoms in MF whereas less contribution arises from Ag-$4d$ states for SF in this energy range. The states at -2.5 eV below Fermi energy are derived from $5d$ and $6s$-states of Hg atom. The top of the valence band is mainly dominated by fulminate group (more contribution from $2p$-states of oxygen atom) in both of the compounds SF and MF while $4d$-states of Ag atom are predominant in case of SF but very less contribution from $5d$-states of Hg atom in case of MF. This implies that there exists a strong hybridization between Ag and C when compared to Hg and C atoms. This does strongly suggest that Ag-C bond has more directional bonding nature over Hg-C bond which indicates that SF is more unstable than MF. Moreover, we also observe $s$, $p$-states of fulminate group and $d$-states of metal atom are dominant in the VB and strong hybridization between Ag/Hg and C, N and O atoms of anionic group which shows the covalent nature in the studied compounds in contrast to the ionic fulminates. The N-O, C-N, Ag-C and Hg-C bonds show less compressibility behavior with increasing pressure (see figure \ref{BLA_MF}), this is due to strong hybridization between (Hg/Ag)-$d$ and $s$, $p$-states of C, N, and O atoms leads to strong covalent character. Furthermore, this can be clearly analyzed from electronic charge density maps which are used for accurate description of chemical bonds. The calculated electron charge density maps along various crystallographic planes of MF using TB-mBJ potential are as shown in figure \ref{Charge}. It shows anisotropic bonding interactions and the charge cloud is distributed within the CNO molecule indicating covalent character as previously reported for SF.\cite{kondal} Overall, the C, N, and O atoms are covalently bonded within CNO group and the metal atom is also covalently bonded with CNO group through C atom in both SF and MF compounds. The presence of covalent bonding in SF and MF makes them more sensitive than the ionic fulminates. Therefore, the heavy metal fulminates can find applications as initiators for secondary explosives due to their instability (high sensitivity) which arises from the structure and bonding nature of the materials.
\section{Optical properties}
Investigation of optical properties for energetic materials is interesting because the knowledge of optical constants is useful for determining decomposition mechanism, laser-augmented combustion and ignition. The optical spectra also allow an estimation of surface reflection losses and spatial distribution of radiation absorption.\cite{Isbell} Moreover, electronic structure calculations could provide an information about the nature and location of inter band transitions in crystals. In our previous work,\cite{kondal} we made a detailed analysis of optical spectra of SF-polymorphs without inclusion of SO. In the present study, we mainly focused on the optical spectra of centro-symmetric orthorhombic structures of SF and MF with and without inclusion of SO interactions. The complex dielectric function $\epsilon(\omega)$ = $\epsilon_1(\omega)$ + $i\epsilon_2(\omega)$ can be used to describe the linear response of the system to electromagnetic radiation which is related to interaction of photons with electrons. The imaginary part of dielectric function $\epsilon_2(\omega)$ is obtained from the momentum matrix elements between the occupied and unoccupied wave functions within selection rules. The orthorhombic symmetry of SF and MF allows three non-zero components of the dielectric tensors along [100], [010] and [001] directions. The calculated real $\epsilon_1(\omega)$ and imaginary $\epsilon_2(\omega)$ parts of dielectric function with and without inclusion of SO are as displayed in figure 2 of supplementary material for SF\cite{support} and in figure \ref{Eps_MF} for MF. The major peaks in $\epsilon_2(\omega)$ of SF are mainly arises due to optical transitions between Ag($4d$) $\rightarrow$ N($2p$) states.\cite{kondal} The prominent peaks in $\epsilon_2(\omega)$ for MF are as follows: the peak at 5.9 eV originates from the transition O($2p$) $\rightarrow$ Hg($s$), the peak at around 8.0 eV arises probably from the transition Hg($6s$) $\rightarrow$ N($p$), the peaks in the energy range 10-16 eV are from the transition Hg($5d$) $\rightarrow$ N/C/O($p$) and finally the peaks around 19.5 eV are due to the transition between Hg($5d$) $\rightarrow$ Hg($p$) states along three crystallographic directions. 

\par The $\epsilon_1(\omega)$ can be derived from the $\epsilon_2(\omega)$ using the Kramer-Kronig relations. The calculated real static dielectric constant along three crystallographic directions with (without) SO are found to be 2.00 (2.34), 5.39 (5.56), 2.64 (2.96) for SF and 1.83 (2.09), 5.11 (5.21), 1.99 (2.26) for MF. Using $\epsilon_1(\omega)$ and $\epsilon_2(\omega)$, one can derive other important optical constants such as refraction, reflectivity, absorption and photo conductivity of the materials. The calculated static refractive indices with (without) SO using the dielectric function n = $\sqrt{\epsilon(0)}$ are given by n$_{100}$ = 1.41 (1.53), n$_{010}$ = 2.32 (2.36), n$_{001}$ = 1.62 (1.72) for SF and n$_{100}$ = 1.35 (1.45), n$_{010}$ = 2.26 (2.28), n$_{001}$ = 1.41 (1.50) for MF. Iqbal et al\cite{iqbal1} proposed that high values for the refractive index suggest that directional bonding might be present in the crystal. The authors also observed high refractive index value for SF over MF when the direction of light was parallel to a-axis. From the calculated refractive indices of both of the compounds, we clearly see that SF has high refractive indices along all the crystallographic directions than MF. Apart from the PDOS, the polarized refractive indices also show that SF has more covalent character when compared to MF which implies that SF is more unstable than MF. Also, the obtained refractive indices are distinct in all three crystallographic directions, which indicates the anisotropy of the SF (see figures 3 of supplementary material)\cite{support} and for MF (see figure \ref{ref_MF} (top)). As illustrated in figure \ref{ref_MF} (bottom), the calculated reflectivity spectra show that the reflectivity starts at around 2 $\%$ and reaches to a maximum reflectivity of 12-14 $\%$ along a- (at around 16 eV)  and c (at around 4 eV for SF and 16 eV for MF)-directions whereas it starts at around 15 $\%$ and reaches to a maximum value of 50 $\%$ at around 7 and 9 eV along b-direction for both of the compounds. This implies that SF and MF has maximum reflectivity along b-direction when compared to a- and c-directions. The calculated absorption spectra is shown in figure 4 of supplementary material for SF\cite{support} and figure \ref{absorp_MF} for MF and absorption starts after the energy 3.30 and 4.82 eV for SF and MF, respectively which is the energy band gap between the VB maximum and Conduction band minimum. The absorption coefficients are found to have order of magnitude $\sim$ 10$^7$m$^{-1}$ which shows that absorption of the compounds lie in the Ultra-Violet (UV) region. Photo conductivity is due to the increase in the number of free carriers when photons are absorbed. The calculated photo conductivity shows a wide photo current response in the absorption region of 3.30-25 eV and 4.82-25 eV as shown in figure 4 of supplementary material and figure \ref{absorp_MF} for SF and MF, respectively. Overall, we observe that inclusion of SO interactions has significant influence on optical properties of the heavy metal energetic SF (see figures 2, 3 and 4 of the supplementary material\cite{support}) and MF salts as shown in figures \ref{Eps_MF},  \ref{ref_MF} and \ref{absorp_MF}. Also, SF and MF show a strong anisotropic and wide range of absorption. This results suggest the possible decomposition of SF/MF into Ag/Hg, CO and N$_2$ under the action of UV light. Therefore, SF/MF decompose under the action of UV light and they may explode due to photochemical decomposition.

\section{Conclusions}
In conclusion, $\emph{ab-initio}$ calculations have been performed to understand the pressure dependent structural and elastic properties of long standing primary explosive, MF. Non-empirical van der Waals density functional methods vdW-DF, optB88-vdW and optB86b-vdW reproduce the experimental volume within $\sim 1.7\%$. Among the non-local correction methods tested, optB88-vdW method works well for the examined compound. MF is found to be softer than the well known primary explosives SF, SA and LA. The lattice constant ${\bf b}$ increases whereas lattice constants ${\bf a, c}$ are decreasing with pressure which shows anisotropic compressibility of MF. The calculated linear compressibility and elastic moduli reveal that MF is more sensitive to detonation along c-axis. The Hg...O non-bonded interactions are responsible for high compressibility of MF along c-axis. The semi-local TB-mBJ potential has been used to calculate the electronic structure and optical properties including SO interactions. The computed electronic structures show that the investigated compounds are indirect band gap insulators. We also noticed that SO is more pronounced for $4d$ and $5d$-states of Ag and Hg atoms of SF and MF, respectively. The nature of chemical bonding is analyzed through the calculated partial density of states and charge density maps. The covalent nature might be the reason for more sensitiveness to external stimuli of heavy metal fulminates when compared to ionic fulminates. The effect of SO coupling on the optical properties is found to be significant for both of the compounds. The most probabilistic electric-dipole transitions are found to occur between Ag($4d$) $\rightarrow$ N($2p$) states for SF whereas O($2p$) $\rightarrow$ Hg($s$), Hg($6s$) $\rightarrow$ N($p$) and Hg($5d$) $\rightarrow$ Hg($p$) states for MF. The calculated absorption coefficients are found to be in the order of 10$^7$m$^{-1}$ which shows that SF and MF are found to decompose under the irradiation of UV light.

\section{Acknowledgments}
Authors would like to thank Defense Research and Development Organization (DRDO) through ACRHEM for the financial support under grant No. DRDO/02/0201/2011/00060:ACREHM-PHASE-II, and the CMSD, University of Hyderabad, for providing computational facilities. NYK would like to acknowledge Prof. M. C. Valsakumar, Department of Physics, IIT Palakkad for his valuable discussions and suggestions. $^*$\emph{Author for Correspondence, E-mail: gvsp@uohyd.ernet.in} 

{}

{\pagestyle{empty}

\begin{table*}[h]
\caption{Calculated ground state lattice parameters (a, b, and c in, $\AA$), volume (V in, $\AA^3$), density ($\rho$ in, gr/cc) of orthorhombic MF using various non-local correction methods. Experimental
data have been taken from Ref. \onlinecite{beck} and the relative errors were given in parentheses with respect to experimental data. Here "-" and "+" represent under- and overestimation of calculated values when compared to the experiments.}
\begin{ruledtabular}
\begin{tabular}{cccccccc}
  Parameter      &  vdW-DF      &   vdW-DF2     &  optB88-vdW  & optB86b-vdW  &  Expt.\cite{beck}  \\  \hline
     a           &  5.518       &    5.559      &   5.451      &    5.447     &  5.470  \\
                 & (+0.9$\%$)   &  (+1.6$\%$)   &  (-0.3$\%$)  &  (-0.4$\%$)  &         \\
     b           &  10.749      &    10.742     &   10.677     &  10.661      &  10.376 \\
                 & (+3.6$\%$)   &  (+3.5$\%$)   &  (+2.9$\%$)  &  (+2.7$\%$)  &         \\
     c           &  7.497       &    7.748      &    7.478     &   7.445      &  7.700  \\
                 & (-2.6$\%$)   &  (+0.6$\%$)   &  (-2.9$\%$)  &  (-3.3$\%$)  &         \\
     V           &  444.67      &    462.67     &  435.22      &   432.33     &  437.03  \\
                 & (+1.7$\%$)   &  (+5.9$\%$)   &  (-0.4$\%$)  &  (-1.1$\%$)  &          \\
   $\rho$        &  4.251       &    4.086      &  4.343       &    4.372     &  4.33    \\
                 & (-1.8$\%$)   &  (-5.6$\%$)   & (+0.3$\%$)   &  (-1.0$\%$)  &          \\
\end{tabular}
\end{ruledtabular}
\label{}
\end{table*}


\begin{table*}[h]
\caption{Calculated single elastic moduli (C$_{ij}$, in GPa), polycrystalline bulk (B$_X$, in GPa) and shear moduli (G$_X$, in GPa) in the Voigt, Reuss, and Hill approximations (X = V, R, H, respectively), Young's modulus (E, in GPa), the longitudinal, transverse, and average sound wave velocities ($v_l$, $v_t$, and $v_m$, in km/s) and Debye temperature ($\theta_D$, in K) of MF using optB88-vdW method.}
\begin{ruledtabular}
\begin{tabular}{cccccc}
    Elastic moduli &  &  Polycrystalline elastic moduli & &  Sound wave velocities  \\ \hline
 C$_{11}$ &  24.7  & B$_V$    &  22.2  & $v_l$       & 2.43  \\  
 C$_{22}$ &  68.2  & G$_V$    &  7.3   & $v_t$       & 1.13  \\ 
 C$_{33}$ &  17.7  & B$_R$    &  14.2  & $v_m$       & 1.27  \\
 C$_{44}$ &  3.6   & G$_R$    &  3.8   & $\theta_D$  & 152  \\
 C$_{55}$ &  8.5   & B$_H$    &  18.2  &  \\
 C$_{66}$ &  2.5   & G$_H$    &  5.6   &  \\
 C$_{12}$ &  8.1   &   E      &  15.1  &  \\
 C$_{13}$ &  10.2  & \\ 
 C$_{23}$ &  26.4  &       \\ 
\end{tabular}
\end{ruledtabular}
\end{table*}


\begin{table*}[h]
\caption{Calculated fractional co-ordinates of SF and MF within PBE-GGA using FP-LAPW method at the experimental lattice constants a = 3.880 $\AA$, b = 10.752 $\AA$, c = 5.804 $\AA$ for SF\cite{Barrick} and 5.47 $\AA$, 10.376 $\AA$, and c = 7.70 $\AA$ for MF.\cite{beck}}
\begin{ruledtabular}
\begin{tabular}{ccccc}
  Compound  &  Atom   &  Wyckoff   &            Present          &      Expt.                 \\  \hline  
SF\cite{Barrick} & Ag &    4$a$    &  (0.0000, 0.0000, 0.0000)   &   (0.0000, 0.0000, 0.0000)    \\
            &   C     &    4$c$    &  (0.0000, 0.1444, 0.2500)   &   (0.0000, 0.1517, 0.2500)    \\
            &   N     &    4$c$    &  (0.0000, 0.2549, 0.2500)   &   (0.0000, 0.2595, 0.2500)    \\
            &   O     &    4$c$    &  (0.0000, 0.3700, 0.2500)   &   (0.0000, 0.3758, 0.2500)    \\
MF\cite{beck} & Hg    &    4$a$    &  (0.0000, 0.0000, 0.0000)   &  (0.0000, 0.0000, 0.0000)    \\
            &   C     &    8$f$    &  (0.0000, 0.8191, 0.0937)   &  (0.0000, 0.8180, 0.0950)    \\
            &   N     &    8$f$    &  (0.0000, 0.7080, 0.1208)   &  (0.0000, 0.7110, 0.1230)    \\  
            &   O     &    8$f$    &  (0.0000, 0.5916, 0.1496)   &  (0.0000, 0.5930, 0.1490)    \\  
\end{tabular}
\end{ruledtabular}
\label{}
\end{table*}

}

\clearpage
\begin{figure*}[h]
\centering
\includegraphics[height = 3.2in, width=6.5in]{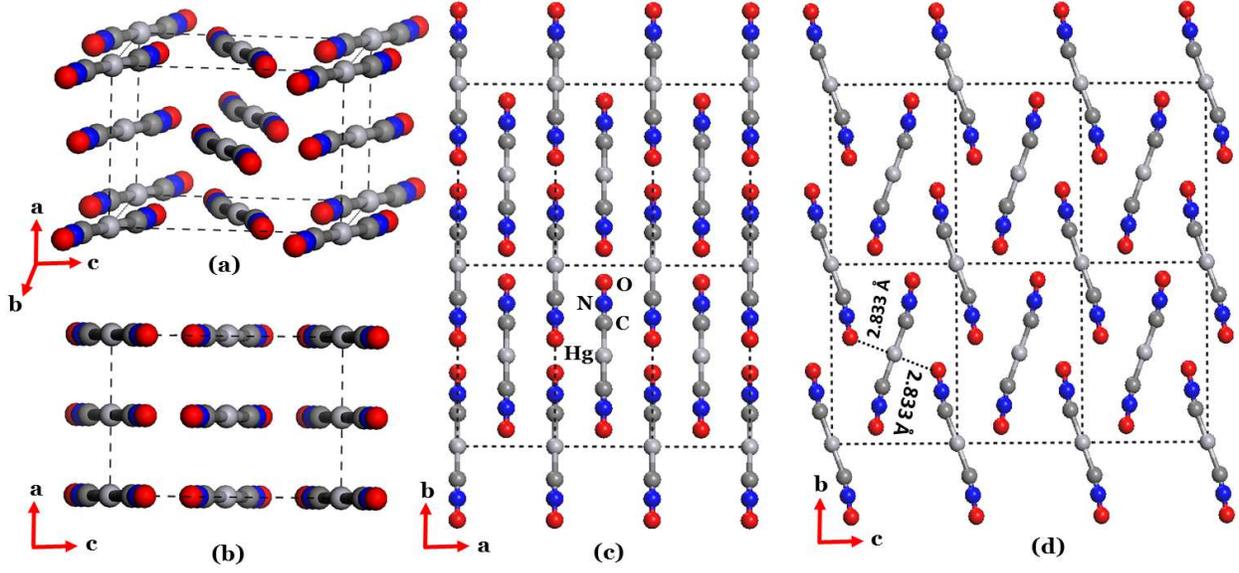}
\caption{(Color online) (a) Unit cell of MF along b-axis, (b, c) Planar layers of MF molecules stacked along a-axis with a distance of $\frac{a}{2}$ = 2.735 $\AA$, and (d) Two equivalent Hg...O = 2.833 $\AA$ non-bonded interactions viewed along c-axis. Light ash, dark ash, blue and red color balls represent mercury, carbon, nitrogen and oxygen atoms, respectively.}
\label{cryst_MF}
\end{figure*}

\begin{figure*}[h]
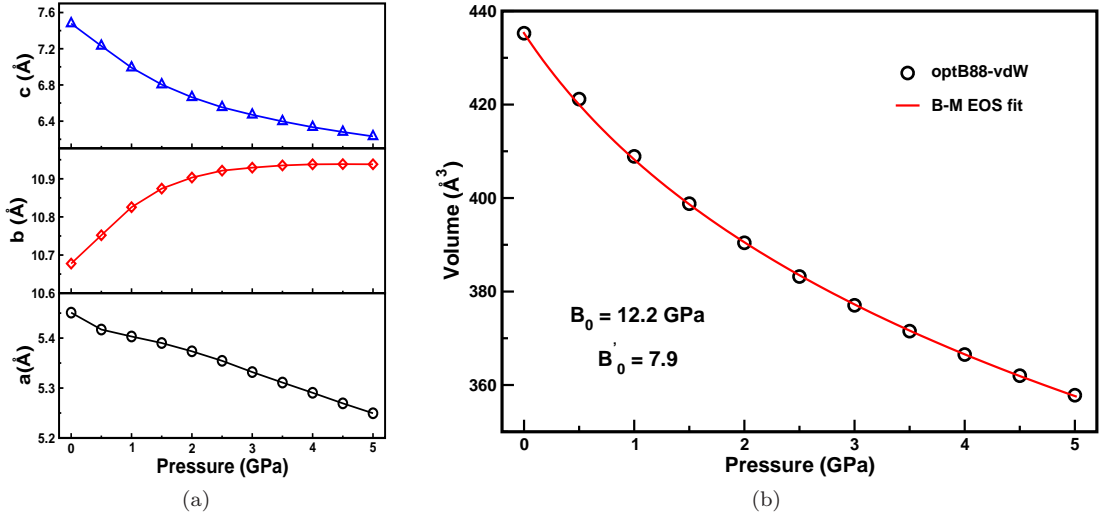

\centering
{\subfigure[]{\includegraphics[height = 2.5in, width=2.0in]{Fig2a.eps}}}  \hspace{0.15 in}
{\subfigure[]{\includegraphics[height = 2.5in, width=3.5in]{Fig2b.eps}}} \hspace{0.15 in}
\caption{(Color online) (a) Calculated lattice constants and (b) volume of MF as a function of pressure using optB88-vdW method.}
\label{abc_MF}
\end{figure*}


\begin{figure*}[h]
\centering
\includegraphics[height = 2.0in, width=6.5in]{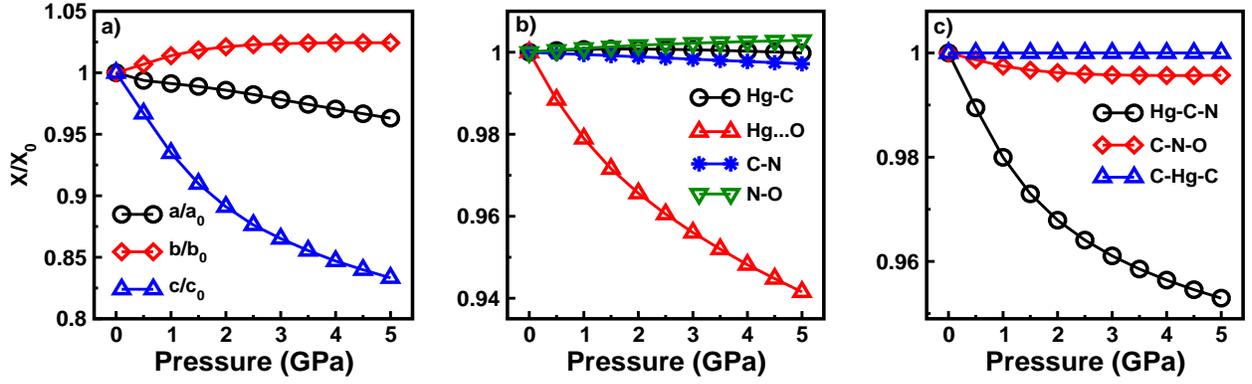}
\caption{(Color online) Calculated normalized (a) lattice constants, (b) bond lengths and (c) angles of MF as a function of pressure using optB88-vdW method. Where X$_0$ and X represent obtained lattice parameters at ambient and as a function of pressure, respectively.}
\label{BLA_MF}
\end{figure*}

\begin{figure*}[h]
\centering
\includegraphics[height = 3.5in, width=5.0in]{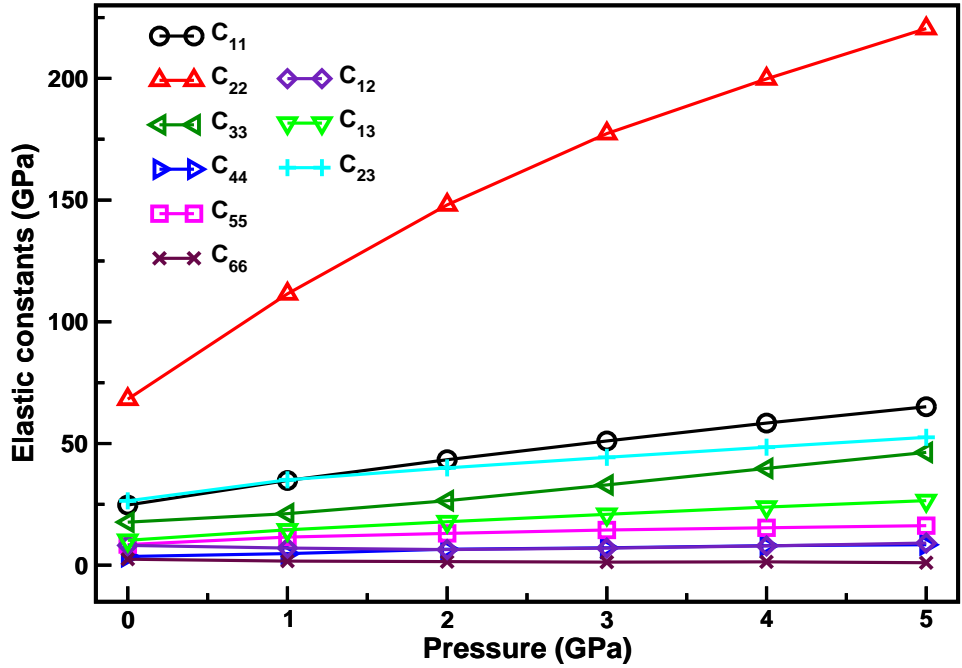}
\caption{(Color online) Calculated elastic constants of MF as a function of pressure using optB88-vdW method.}
\label{Cij_MF}
\end{figure*}

\begin{figure*}[h]
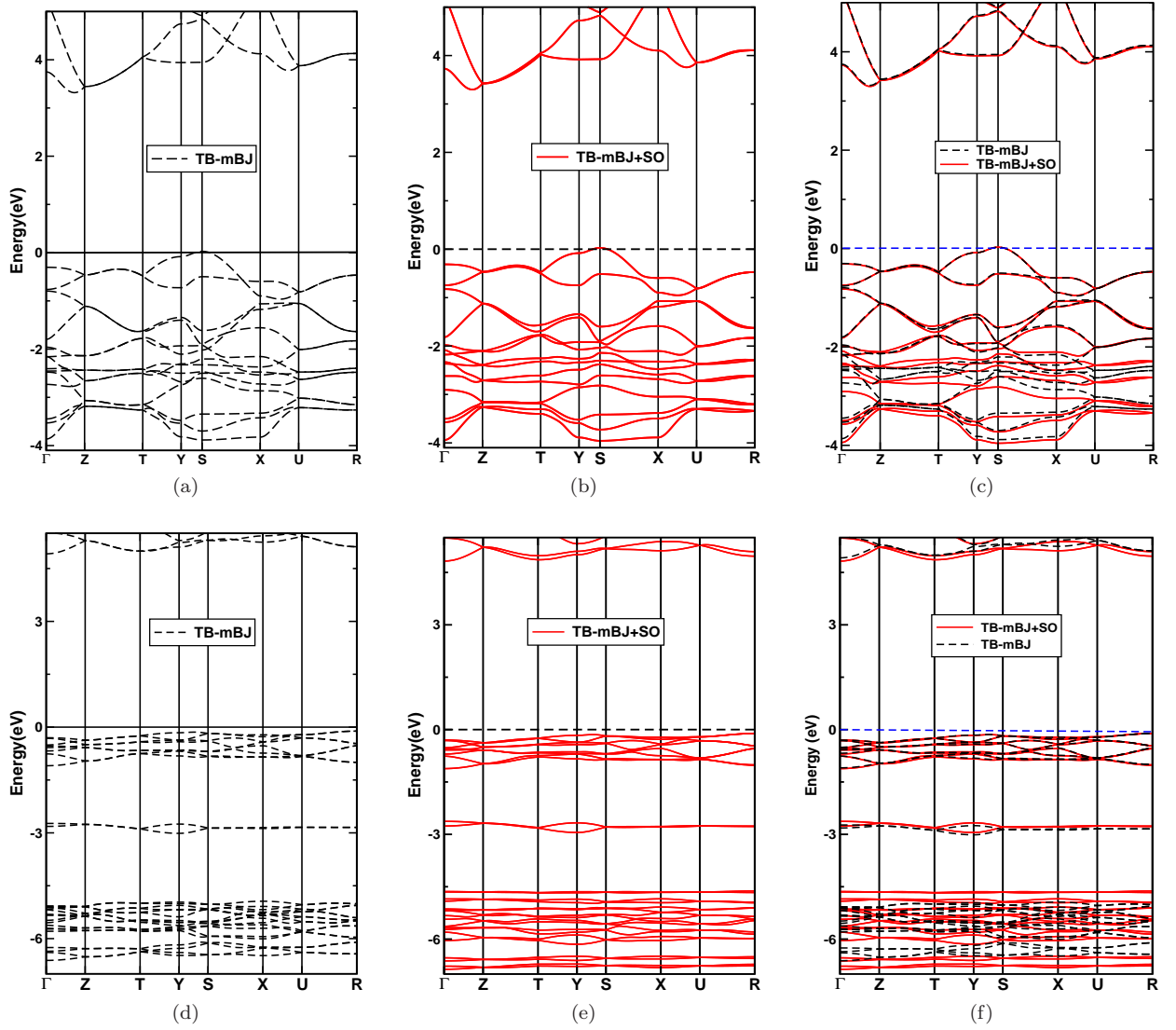

\centering
{\subfigure[]{\includegraphics[height = 2.7in, width=2.0in]{Fig5a.eps}}}     \hspace{0.15 in}
{\subfigure[]{\includegraphics[height = 2.6in, width=2.0in]{Fig5b.eps}}}  \hspace{0.15 in}
{\subfigure[]{\includegraphics[height = 2.75in, width=2.0in]{Fig5c.eps}}}  
{\subfigure[]{\includegraphics[height = 2.6in, width=2.0in]{Fig5d.eps}}}     \hspace{0.15 in}
{\subfigure[]{\includegraphics[height = 2.7in, width=2.0in]{Fig5e.eps}}}  \hspace{0.15 in}
{\subfigure[]{\includegraphics[height = 2.7in, width=2.0in]{Fig5f.eps}}}       
\caption{(Color online) Calculated electronic band structures of (a, b, c) SF (top) and (c, d, e) MF (bottom) without (black dotted lines) and with (solid red lines) inclusion of SO coupling using the TB-mBJ potential at the experimental lattice constants.\cite{Barrick,beck}}
\label{BS}
\end{figure*}

\begin{figure*}[h]
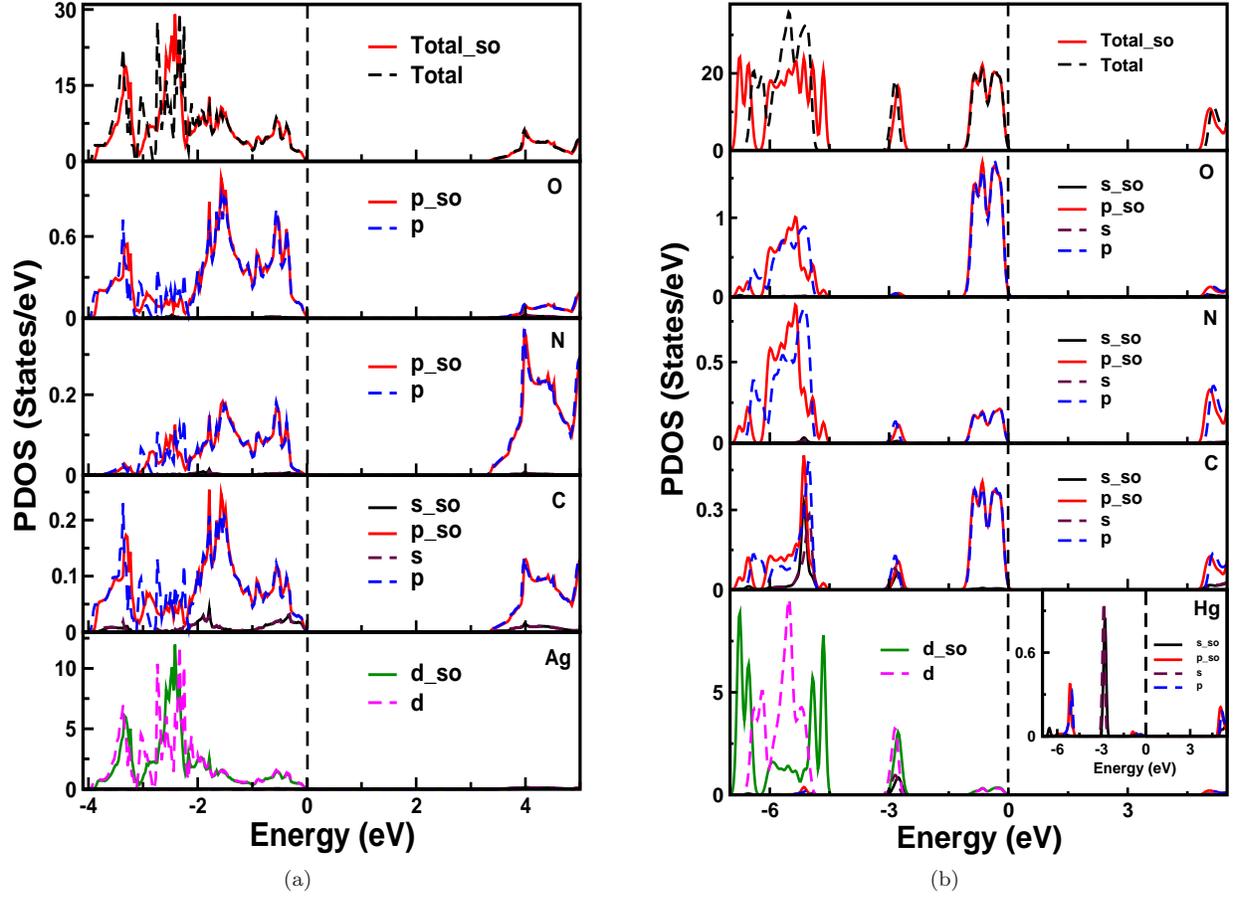

\centering
{\subfigure[]{\includegraphics[height = 4.5in, width=3.0in]{Fig6a.eps}}} \hspace{0.3 in}
{\subfigure[]{\includegraphics[height = 4.5in, width=3.0in]{Fig6b.eps}}}
\caption{(Color online) Calculated total and partial density of states of SF (left) and MF (right) with and without inclusion of SO interactions using the TB-mBJ potential at the experimental lattice constants.\cite{Barrick,beck}}
\label{DOS}
\end{figure*}

\begin{figure*}[h]
\centering
\includegraphics[height = 3.2in, width=6.8in]{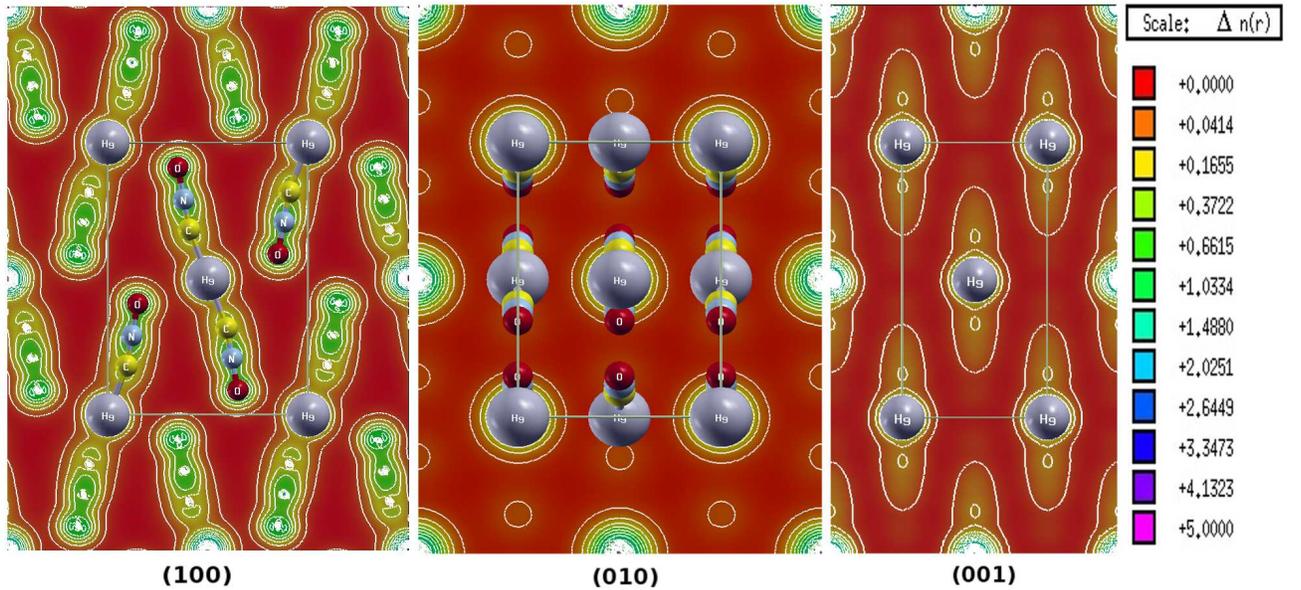}
\caption{(Color online) Calculated electronic charge densities of MF along crystallographic (100), (010), and (001) planes.}
\label{Charge}
\end{figure*}

\begin{figure*}[h]
\centering
\includegraphics[height = 3.0in, width=6.2in]{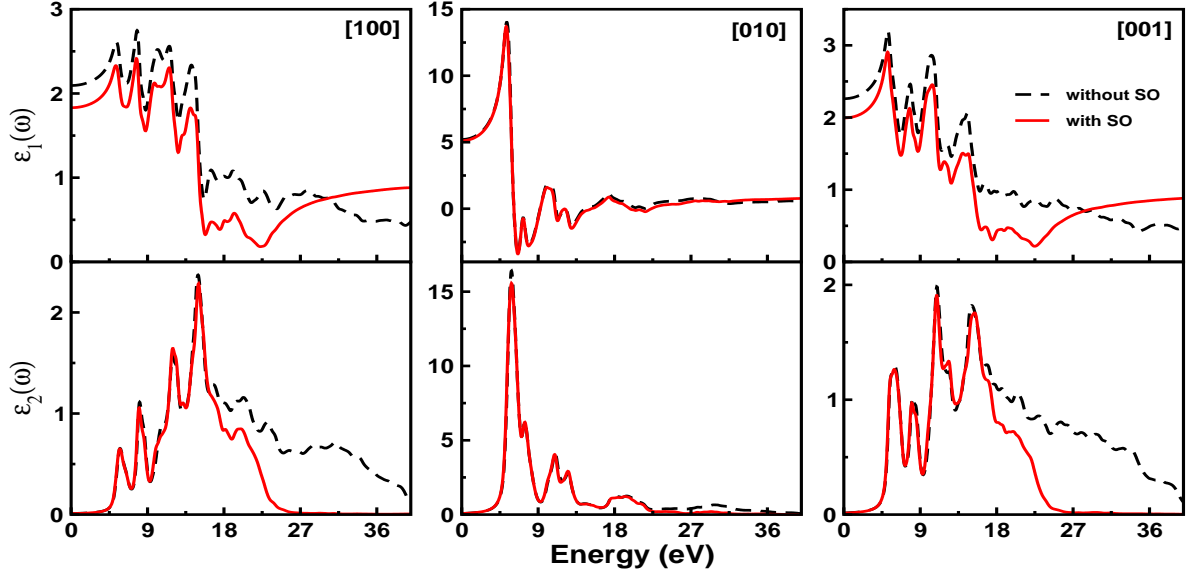}
\caption{(Color online) Calculated real ($\epsilon_1(\omega)$) and imaginary ($\epsilon_2(\omega)$) parts of complex dielectric function of MF with (solid red lines) and without (dotted black lines) inclusion of SO interactions using the TB-mBJ potential at the experimental lattice constants.\cite{beck}}
\label{Eps_MF}
\end{figure*}

\begin{figure*}[h]
\centering
\includegraphics[height = 3.0in, width=6.2in]{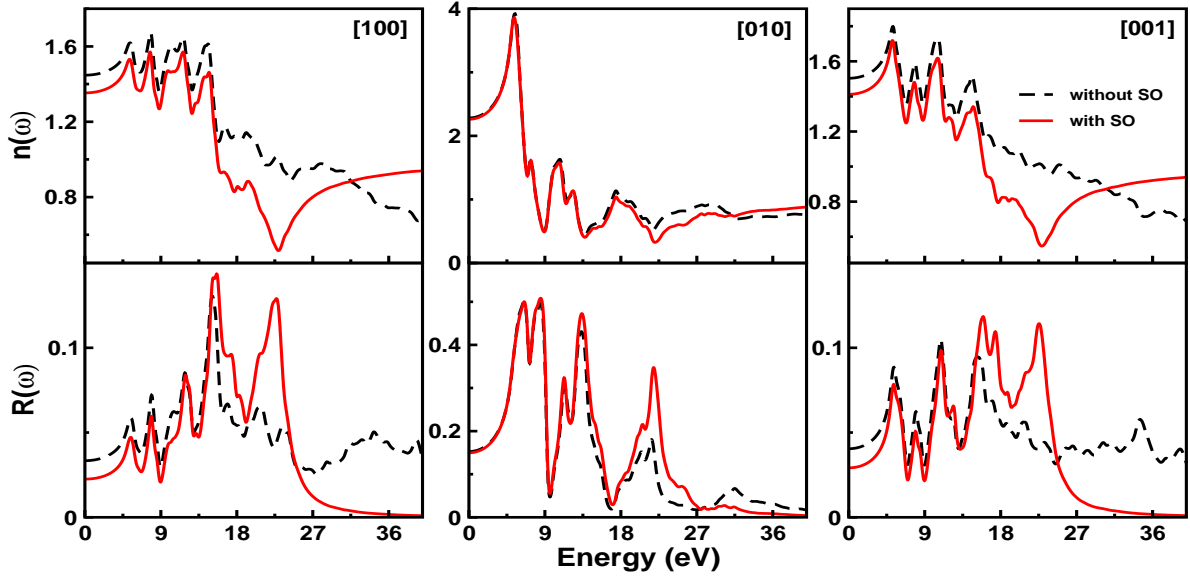}
\caption{(Color online) Calculated refraction (n($\omega$)) and reflectivity (R($\omega$)) spectra of MF with (solid red lines) and without (dotted black lines) inclusion of SO interactions using the TB-mBJ potential at the experimental lattice constants.\cite{beck}}
\label{ref_MF}
\end{figure*}

\begin{figure*}[h]
\centering
\includegraphics[height = 3.0in, width=6.2in]{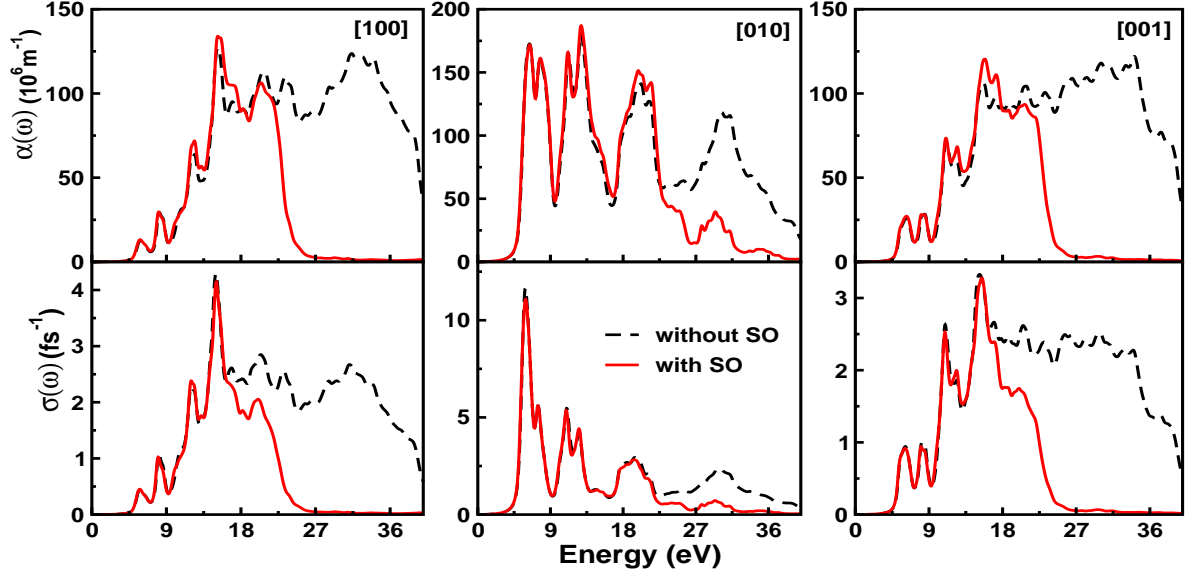}
\caption{(Color online) Calculated absorption ($\alpha(\omega$)) and photo conductivity ($\sigma(\omega$)) spectra of MF with (solid red lines) and without (dotted black lines) inclusion of SO interactions using the TB-mBJ potential at the experimental lattice constants.\cite{beck}}
\label{absorp_MF}
\end{figure*}

\end{document}